\documentclass[a4paper]{jpconf}

\usepackage{hyperref}

\newcommand{\footlabel}[2]{%
    \addtocounter{footnote}{1}%
    \footnotetext[\thefootnote]{%
        \addtocounter{footnote}{-1}%
        \refstepcounter{footnote}\label{#1}%
        #2%
    }%
    $^{\ref{#1}}$%
}

\newcommand{\footref}[1]{%
    $^{\ref{#1}}$%
}

\usepackage{graphicx}
\usepackage{lineno}
\begin{document}
\title{Elliptic flow of heavy-flavour decay electrons in Pb-Pb collision at $\sqrt{s_\mathrm{NN}}$ = 2.76 TeV with ALICE}

\author{Andrea Dubla for the ALICE Collaboration}

\address{Department of Physics, Utrecht University, Princetonplein 5, 3584 CC Utrecht, The Netherlands}

\ead{andrea.dubla@cern.ch}


\begin{abstract}
We present the elliptic flow $v_{2}$ of electrons from the semi-leptonic decays of heavy-flavour hadrons at mid-rapidity for semi-central (20-40\% centrality) Pb-Pb collisions at $\sqrt{s_\mathrm{NN}}$ = 2.76 TeV measured  with the ALICE detector. 
A positive $v_{2}$ is observed at low $p_\mathrm{T}$ in semi-central collisions. Comparisons with previous measurements obtained at lower collision energy at RHIC in Au-Au collisions at $\sqrt{s_\mathrm{NN}}$ = 0.2 TeV and with theoretical models are also presented.
%
\end{abstract}

\section{Introduction}
Heavy-quarks, i.e. charm and beauty, are mainly produced in hard scattering processes in the early stages of high energy nucleus-nucleus collisions and they experience the full evolution of the medium formed in such collisions, the Quark-Gluon Plasma (QGP). Therefore, heavy-flavour hadrons are well suited probes to investigate the properties of the medium.
Due to their large masses, charm and beauty quarks are expected to lose less energy than light quarks and
gluons when traversing the medium due to the dead-cone effect \cite{deadcone} and also to have a longer relaxation time  \cite{relaxation}.
%
In this contribution we present the measurement of elliptic flow $v_{2}$ of electrons from the semi-leptonic decays of heavy-flavour hadrons in semi-central (20-40\% centrality) Pb-Pb collisions at $\sqrt{s_\mathrm{NN}}$ = 2.76 TeV with ALICE. The elliptic flow is the second Fourier coefficient of the azimuthal distribution of particle momenta in the transverse plane with respect to the reaction plane.
At low $p_\mathrm{T}$, the $v_{2}$ of the electrons from heavy-flavour
hadron decays is sensitive to the degree of thermalization of charm and
beauty quarks in the deconfined medium. At higher $p_\mathrm{T}$, the measurement of $v_{2}$
carries information on the path length dependence of in-medium parton energy
loss.
The analysis was carried out with different methods that have different sensitivity to non-flow contributions, namely event plane \cite{EP}, scalar product  \cite{SP} and Q-Cumulants  \cite{qc}.
We compare the experimental result to several theoretical models and with measurements from Au-Au collisions at $\sqrt{s_\mathrm{NN}}$ = 0.2 TeV at RHIC \cite{phenix}.


\section{Data analysis}
The data analyzed were recorded with the ALICE detector \cite{alicedetector} during the 2010 and 2011 LHC runs with Pb-Pb collisions at $\sqrt{s_\mathrm{NN}}$ = 2.76 TeV.
The VZERO detector provides the MB trigger, centrality determination, which is used online to enhance the sample of central and semi-central events, and event plane determination.
In addition to the sample from the semi-central trigger (8$\times$$10^6$ events in the centrality range 20-40\%), a sample of 1.3$\times$$10^6$ events in 20-40\% centrality triggered by the Electro-Magnetic Calorimeter (EMCal) is used. 
Electron candidates are selected and identified using different approaches. 
In the low $p_\mathrm{T}$ region ($p_\mathrm{T}$ $<$ 6 GeV/$c$) information from the TPC (Time Projection Chamber) and the TOF (Time Of Flight) detectors is used, while the EMCal contributes to electron identification for 1.5 $<$ $p_\mathrm{T}$ $<$ 13 GeV/$c$.
Particle identification in the TPC is based on the measurement of the specific energy loss dE/dx in the detector gas, while in the TOF the charged particles are separated via the measurement of their time of flight from the interaction point to the detector.
Electron identification in the EMCal is based on the measurement of the E/p ratio, where E is the energy of the EMCal cluster that was matched to the prolongation of the track with momentum $p$ reconstructed in the TPC and ITS detectors. Electrons deposit their total energy in the EMCal and due to their small mass the E/p ratio should be equal to unity. 

The sample of identified electron candidates is mainly composed of two distinct sources: electrons from the semi-leptonic decays of hadrons containing a
charm or beauty quark, and photonic electrons, originating mainly from Dalitz decays of $\pi^{0}$ and $\eta$ mesons and from photon conversion.
In order to extract the heavy-flavour signal, the background contributions have to be subtracted from the inclusive electron spectra and $v_{2}$. In this analysis this task is accomplished using two different methods. In the first method  a cocktail of electron spectra from background sources is calculated using a Monte Carlo event generator of hadron decays \cite{cockt}.
In the second method the electron background is reconstructed using the invariant mass method. 
The latter makes use of the fact that an electron from
a photonic source is always accompanied by a positron, forming an
electron-positron pair with small invariant mass. In the current analysis
pairs with a mass below 70 MeV/$c^{2}$ are tagged as photonic.

\begin{figure}[ht]
\begin{minipage}[t]{0.55\linewidth}
\centering
   \includegraphics[scale=0.35]{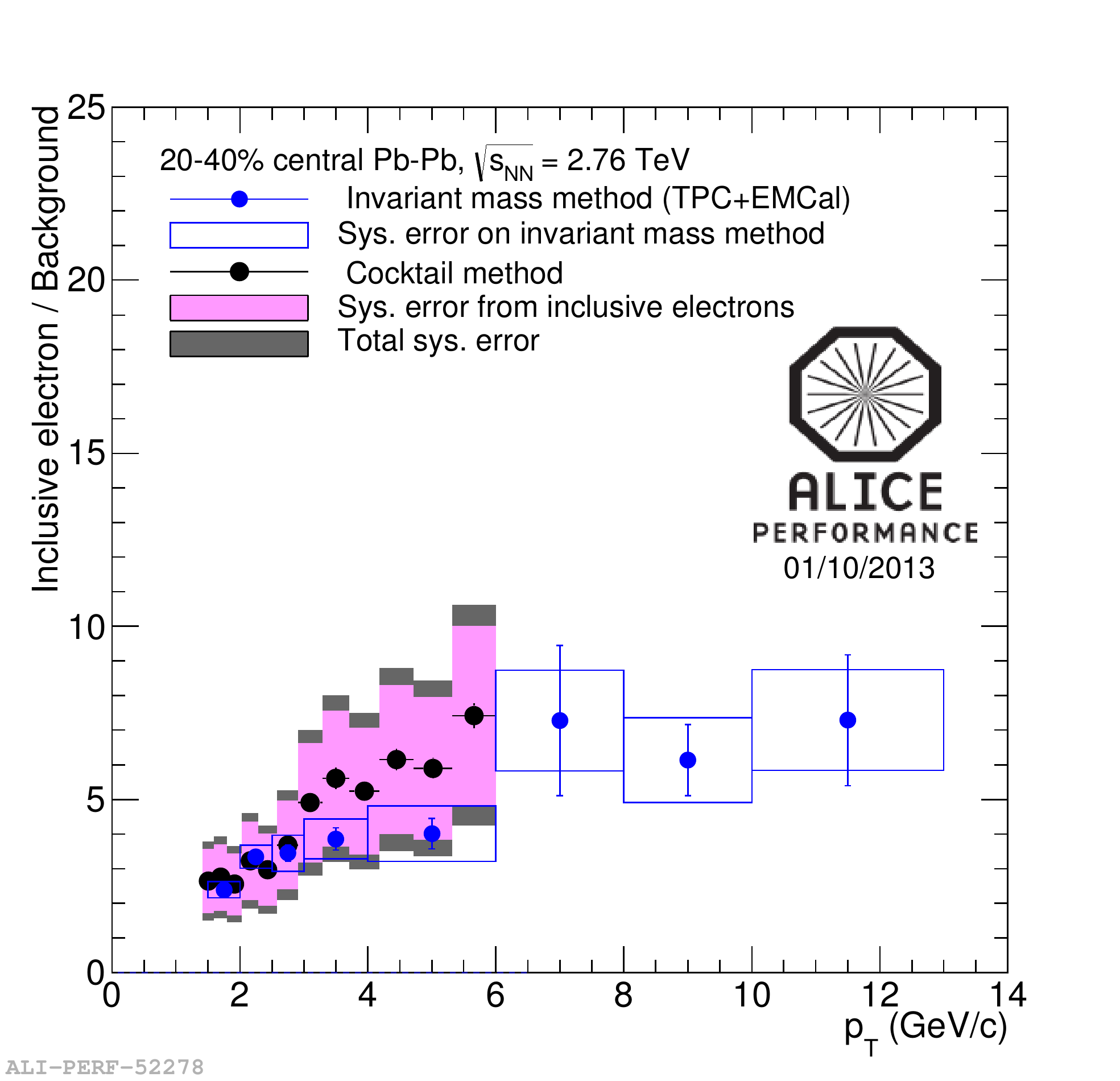}
   \caption{Inclusive over background electron ratio estimated with the cocktail method (black marker) and with the invariant mass method (blue marker).}
    \label{rsb}
\end{minipage}
\hspace{0.3cm}
\begin{minipage}[t]{0.45\linewidth}
\centering
  \includegraphics[scale=0.35]{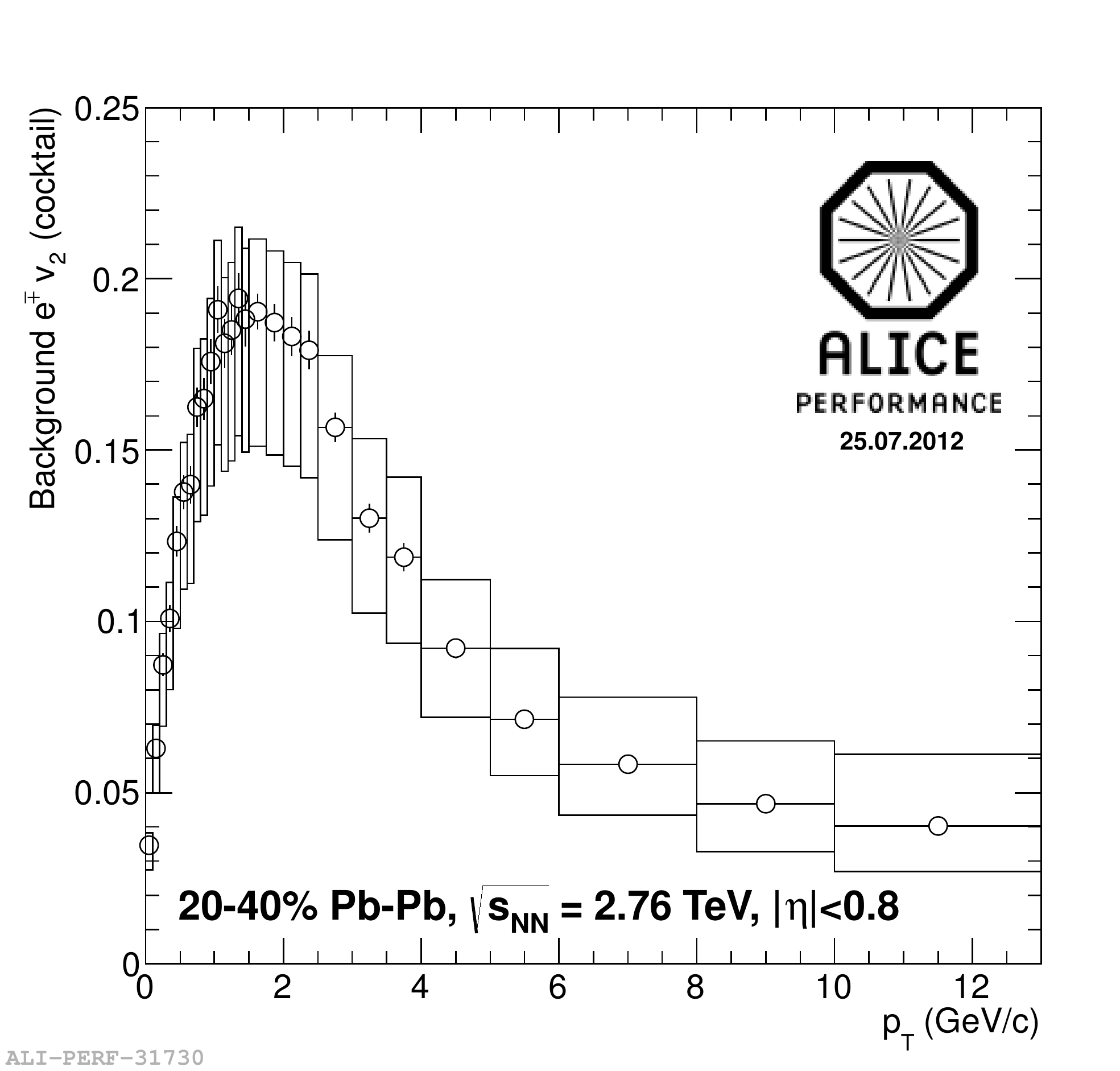}
   \caption{Background electron $v_{2}$ estimated with the cocktail method for Pb-Pb collisions in the 20-40\% centrality range.}
 \label{cock}
\end{minipage}
\end{figure}

The heavy-flavour electron $v_{2}^\mathrm{HFe}$ was then obtained as:

\begin{equation}
v_{2}^\mathrm{HFe} = \frac{(1 + R_\mathrm{SB}) v_{2}^{e}  - v_{2}^{\gamma_{e}}}{R_\mathrm{SB}} 
\end{equation} 

where $v_{2}^{e}$ is the inclusive electron $v_{2}$, (1 + $R_\mathrm{SB}$) is the ratio of inclusive electrons to the estimated background electrons and $v_{2}^{\gamma_{e}}$ is the $v_{2}$ coefficient for the background sources estimated from the cocktail. The (1 + $R_\mathrm{SB}$) was estimated with both methods, cocktail and invariant mass method as a function of $p_\mathrm{T}$. The ratio of the $p_\mathrm{T}$ spectra comparing the two different methods is shown in  in Fig. \ref{rsb}\footlabel{rom}{Fig. \ref{rsb} and Fig. \ref{fvqef} are updated versions of the plots shown in the conference. The $v_{2}$ results in the transverse momentum intervals 8 $< p_\mathrm{T} <$ 10 GeV/c and 10 $< p_\mathrm{T} <$ 13 GeV/c were added for the second order cumulants and for the scalar product methods.}
and a good agreement is observed within statistical and systematic uncertainties. The $v_{2}^{\gamma_{e}}$ estimated for the background with the cocktail is reported in Fig. \ref{cock}.

\section{Results and conclusions}

\begin{figure}[ht]
\begin{minipage}[t]{0.445\linewidth}
\centering
   \includegraphics[scale=0.39]{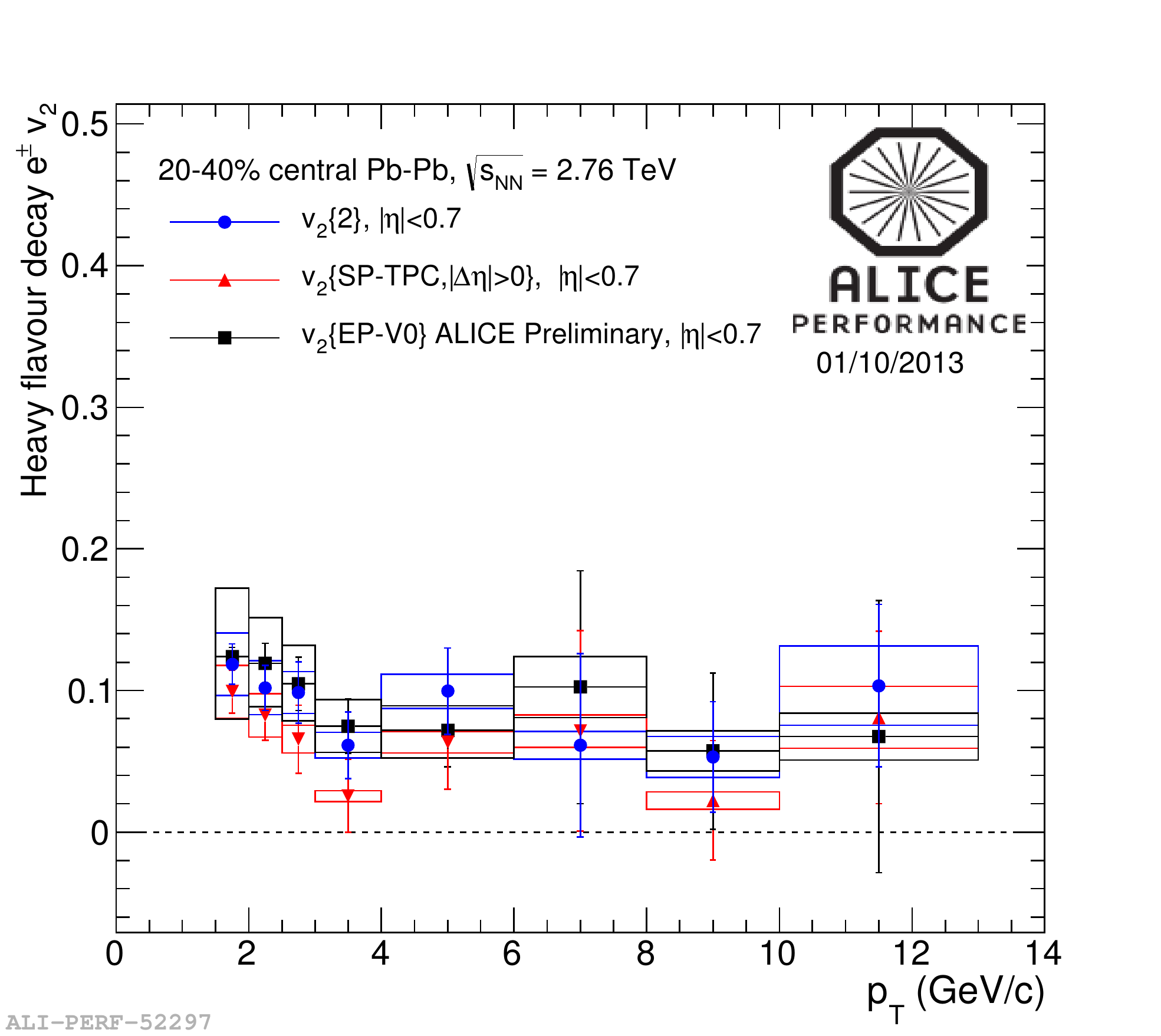}
   \caption{Comparison of heavy flavor decay electron $v_{2}$ extracted with the second order cumulants (blue marker),  scalar product (red marker) and event plane (black marker) methods. }
    \label{fvqef}
\end{minipage}
\hspace{0.3cm}
\begin{minipage}[t]{0.555\linewidth}
\centering
  \includegraphics[scale=0.35]{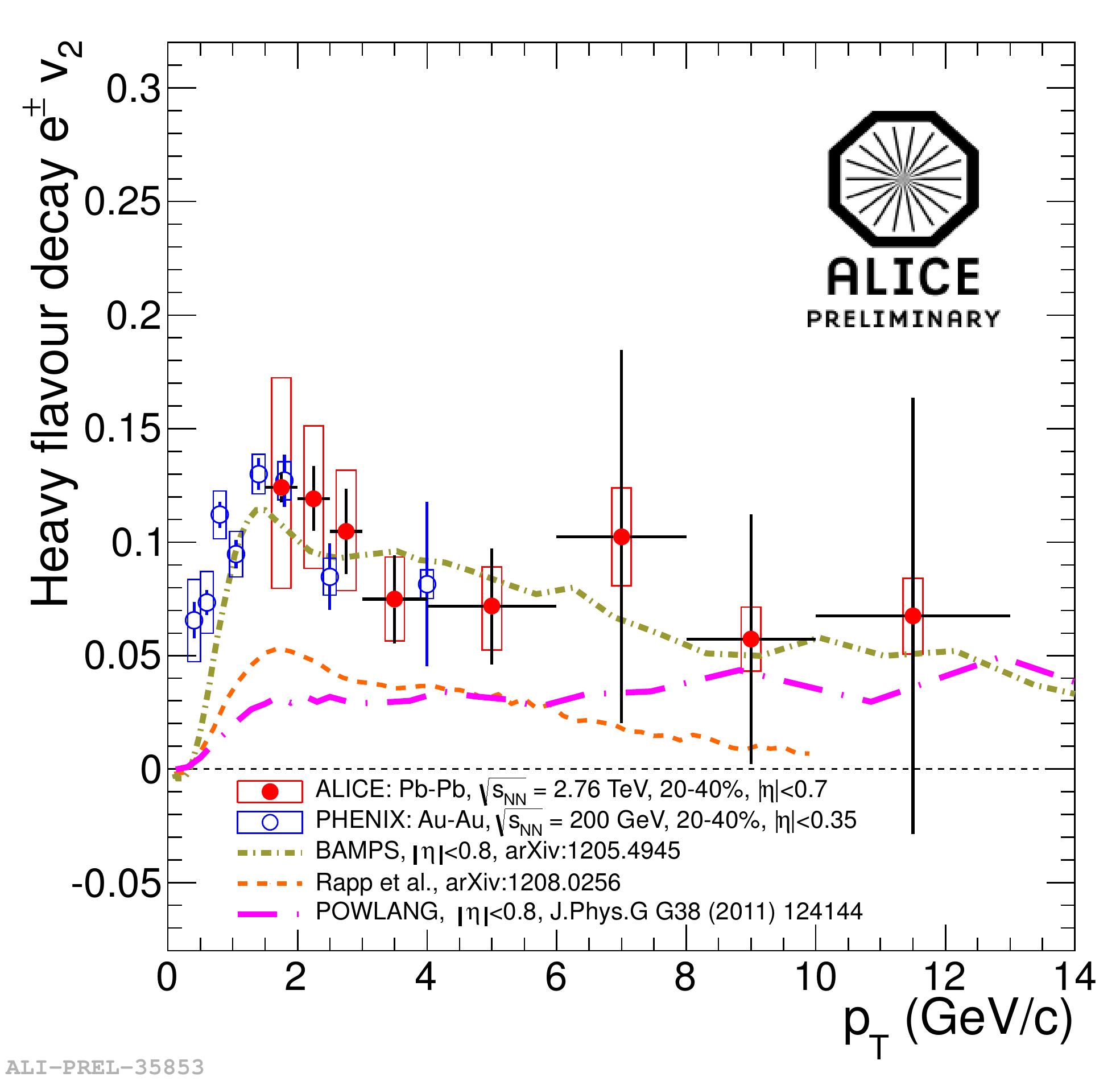}
   \caption{Heavy-flavour decay electrons $v_{2}$ measured with the event plane method in 20-40\% Pb-Pb collisions at $\sqrt{s_\mathrm{NN}}$ = 2.76 TeV (red marker) compared to different models specific for LHC and with PHENIX results in Au-Au collisions at $\sqrt{s_\mathrm{NN}}$ = 0.2 TeV (blue marker). }
   \label{rtht}
\end{minipage}
\end{figure}

The transverse momentum dependence of $v_{2}$ of heavy-flavour decay electrons is shown in Fig. \ref{fvqef}\footref{rom} for the different methods used in the analysis, which are found to be consistent within the uncertainties. 
We observe  positive $v_{2}$ at low $p_\mathrm{T}$ in semi-central collisions.
The positive $v_{2}$ has a contribution from heavy-quark collective flow at low $p_\mathrm{T}$ and may be due to the path-length dependence of the energy loss mechanism. 
In Fig. \ref{rtht} the experimental results are compared with measurements by PHENIX in Au-Au collisions at RHIC at $\sqrt{s_\mathrm{NN}}$ = 0.2 TeV \cite{phenix} and with various transport model calculations.
The  $v_{2}$ measured at the LHC is similar in magnitude to that measured by PHENIX at lower collision energy.
The model calculations shown in Fig. \ref{rtht} incorporate parton transport based on the Boltzmann approach that includes the coupling of the quarks with an expanding medium via collisional and radiative processes (BAMPS) \cite{bamps}; heavy-quarks transport with in-medium resonance scattering and coalescence (Rapp et. al) \cite{rap}; and heavy-quarks transport (Langevin equation) with collisional energy loss (POWLANG) \cite{powlang}.
The BAMPS calculation is in good agreement with the $v_{2}$ measurement while the calculations by Rapp et al. and POWLANG underestimate the $v_{2}$. 

\section{References}



\begin{thebibliography}{9}
\bibitem{deadcone}
Dokshitzer Y L and Kharzeev D E  arXiv:hep-ph/0106202 
\bibitem{relaxation}
Moore G D and Teaney D 
  \textit{Phys. Rev.} C {\bf 71} (2005) 064904
\bibitem{EP}
Poskanzer A M and Voloshin S A  \textit{Phys. Rev.} C \textbf{58}, 1671
\bibitem{SP}
C. Adler et al. [STAR Coll.] \textit{Phys. Rev.} C \textbf{66}, 03 4904 (2002)
\bibitem{qc}
Bilandzic A, Snellings R, Voloshin S  arxiv.org/abs/1010.0233
\bibitem{cockt}
B. Abelev et al. [ALICE Coll.] arXiv:1205.5423
\bibitem{phenix}
A.Adare et al. [PHENIXColl.] \textit{Phys.Rev.} C \textbf{84}, 04 4905 (2011)
\bibitem{alicedetector}
K. Aamodt et al. [ALICE Coll.] \textit{JINST} 3, 2008 
\bibitem{bamps}
Uphoff J, Fochler O, Xu Z and Greiner C  arXiv:1205.4945
\bibitem{rap}
He M, Fries R J and Rapp R  arXiv:1106.6006v1
\bibitem{powlang}
Monteno M et al. \textit{J. Phys.} G \textbf{38} 124144


\end{thebibliography}
\end{document}